# A correlation of structural changes with nanomechanical properties in TiN-AlN multilayer films

**Nidhin George Mathews[1,2], Aidan A. Taylor[3], Johannes Zechner[3], Helmut Riedl[4], Paul H. Mayrhofer[4], Johann Michler[3,5], Vipin Chawla[3,6*], Gaurav Mohanty[1*]**

[1] Materials Science and Environmental Engineering, Tampere University, 33014 Tampere, Finland

[2] VTT Technical Research Center of Finland, 02044 Espoo, Finland

[3] Mechanics of Materials and Nanostructures Laboratory, Empa- Swiss Federal Laboratories for Materials Science and Technology, CH-3602 Thun, Switzerland

[4] Institute of Materials Science and Technology, TU Wien, A-10640 Vienna, Austria

[5] École Polytechnique Fédérale de Lausanne (EPFL), 1015 Lausanne, Switzerland

[6] Indian Institute of Technology Roorkee, Roorkee, Uttarakhand 247667, India

***Corresponding authors:** vipin.phy@gmail.com; gaurav.mohanty@tuni.fi

**Abstract**

The present work investigates the changes in overall nanomechanical properties of reactively sputtered TiN-AlN multilayer films arising due to phase transformation in the AlN layers. Multilayered TiN-AlN films were sputter deposited with constant TiN layer thickness of 5 nm while the AlN layer thickness varied between 1-5 nm. The AlN underwent a phase transition from cubic rock salt to hexagonal wurtzite above 3 nm thickness due to the lattice strains. The hardness and indentation modulus of the multilayers decreased with increasing AlN film thickness, up to 3 nm, due to increased volume fraction of softer AlN layer and then stabilized for 4 nm and 5 nm thickness films. Micropillar compression of these multilayers showed a transition from columnar brittle to partially ductile failure associated with crack deflection with increasing AlN film thickness. Interestingly, nanoindentation scratch resistance of 3 nm AlN multilayer was observed to be superior compared to all other films. The crack propagation behavior in scratching showed increased microcracking tendency towards higher AlN film thickness. This shows that cubic to hexagonal transformation in AlN is beneficial for improving the damage tolerance of the multilayer system.

## 1. Introduction

In comparison to single layer films, multilayer films offer improved mechanical performance due to the possibility to match the composite layers based on their individual properties [1–4]. For very thin layers (< 50 nm), the deformation mechanisms and the phase stability might change because of size effects and epitaxial growth. Multilayered films can overcome many performance short comings of single layer films, due to changes in deformation mechanisms like dislocation confinement in layers, etc. [5,6]. The deposition of several thin layers with different mechanical properties allows the designer to meet complex requirements like hardness, toughness, wear resistance and adhesion. Lastly, interfacial areas favor stress relaxation and prevent the propagation of cracks [7]. A long list of properties has to be met like high thermal and chemical stability, immiscibility of the individual layers (to prevent interdiffusion as well as interface transition), form coherent surfaces, high hardness and toughness, good wear/abrasion resistance, high adhesion and good corrosion resistance for using multilayer films in tribological applications such as cutting and machining tool industry, and as protective films for turbine blades and engine parts. [1–4,8–10].

Unlike many nitrides, Titanium Nitride (TiN) and Aluminum Nitride (AlN) are immiscible, so sharp and stable interfaces are expected to form, promoting large hardness enhancement within a wide range of temperatures, along with good oxidation resistance [9,11,12]. The TiN-AlN multilayer system has been chosen for the investigation to understand the fundamental aspects of phase stability and for exploring the strengthening mechanisms in such composite structures. The TiN films show higher hardness and thermal stability in comparison to AlN, while AlN possesses elevated thermal conductivity and bending strength against TiN [13]. In previous studies, researchers have reported that hardness enhancement and toughness improvement depends on the number and quality of interfaces [7,8,14]. This improvement in mechanical properties has been related to the phase transition from hexagonal wurtzite AlN (w-AlN) to cubic AlN (c-AlN) or to the pseudomorphic stabilization of the non-equilibrium rock salt type (B1) AlN [14]. Auger et al. [12] studied TiN-AlN bilayer and multilayer film systems for total thicknesses of 800 nm and 50 nm varying the individual layer thickness and number of interfaces. They observed fine granular structure with the grain size restricted by the layer thickness for the multilayer form with 5 nm bilayers. No well-resolved peaks were observed in X-ray diffraction pattern of 5-5 nm sample due to small grain sizes. However, no hardness values were obtained in 5 nm multilayers due to the reason that the film being too thin. In other samples, the hardness enhancement was observed with increasing number of interfaces due to the dislocation blocking at the interfaces. Setoyama et al. [8]

confirmed the microstructural stability of TiN-AlN multilayers up to 1000 °C in inert environments but the TiN/AlN superlattice structure was observed to change substantially after annealing at 1100 °C. Moreover, this system shows the properties of $Ti_{1-x}Al_xN$ protective films i.e. relatively high hardness and good oxidation resistance. Kim et al. [11] and Madan et al. [15] discuss the transformation of c-AlN to w-AlN on increasing layer thickness above a critical value. Several mechanisms like structure barrier, coherency stresses, misfit dislocations and modulus mismatch have been proposed to explain greater strength of multilayers as reviewed by Clemens et al. [16] and Geisler et al. [17].

Other publications on superlattice and multilayer structures of TiN-AlN films focused on the mechanical characterization especially nanoindentation, scratch and wear testing to analyze and obtain variety of results [7,8,11,12,18–25]. The majority of the published literatures considers the films with equal thicknesses of TiN and AlN layers for periods ranging from 0.8 nm – 1150 nm [7,8,11,12,18,19,22]. The rationale for this study is to specifically investigate the changes in mechanical properties of the TiN-AlN multilayered system associated with the cubic to wurtzite phase transformation, by varying the thickness of AlN layers while keeping the thickness of TiN constant. There is no literature available on the micro compression study of TiN-AlN multilayer films which is a technique to study fracture behavior of the system in terms of coherence and non-coherence between TiN and AlN layers.

Therefore, in the current work, the approach of *in situ* scanning electron microscopy (SEM) micro-compression tests were conducted to evaluate the deformation and fracture mechanisms in TiN-AlN multilayer films. TiN layer thickness was kept constant at 5 nm and the AlN layer thickness was varied in 1 nm steps from 1 to 5 nm so that the influence of two different phases of AlN i.e. cubic and wurtzite (cubic phase transform into wurtzite beyond 2 to 3 nm layer thickness) on the fracture behavior is investigated in detail. Apart from this, nanoindentation and scratch tests have been conducted on the samples to get a broad picture of the hardness and modulus of the TiN-AlN multilayer system with respect to AlN layer thickness and phase change associated with it. Nanoindentation experiments performed in the films determine the hardness and modulus, whereas micropillar compression on these films determines the yield stress for failure of these films in uniaxial compressive stress state, and scratch tests assess the adhesion, fracture resistance and wear behavior of these films.

## 2. Experimental Details

### *2.1. Fabrication of TiN-AlN multilayer films*

TiN-AlN multilayer films were deposited on Si (100) substrates in an AJA Orion 5 magnetron sputtering system. Direct current (dc) sputtering was used to sputter titanium (99.99 % pure, 76.2 mm diameter × 3 mm thick) and aluminum (99.99 % pure, 76.2 mm diameter × 3 mm thick) targets onto the Si (100) substrates with controlled Ar and $N_2$ gas flow rates. A computer-controlled shutter system was used for reliable and reproducible deposition of alternating TiN and AlN layers in the sputtered film. Prior to the sputtering depositions, the chamber was evacuated to a base pressure better than $1\times10^{-4}$ Pa using a turbo molecular pump backed by a rotary pump. The substrates were cleaned before the deposition using a two-step procedure: (a) ultrasonication for 15 minutes in an acetone bath, followed by (b) rf-etching for 5 mins in an Ar glow discharge in the deposition chamber, at pressure of 3 Pa. Both the targets were pre-sputtered for 10 minutes to remove surface contaminants, with the substrate shutter shielding the Si wafer. The multilayer arrangement was kept similar for all depositions: TiN as the starting layer at the substrate-film interface and as the top layer at the end of deposition. The sputtering parameters for TiN-AlN multilayer films are tabulated in **Table 1**. The total thickness of all the deposited films was ~1.4 µm and the substrate was rotated during sputtering to ensure uniform film deposition.

***Table 1:** Sputtering parameters for TiN-AlN multilayer films*

| **Target** | Ti and Al |
|---|---|
| **Base pressure** | < $1\times10^{-4}$ Pa |
| **Gas used** | Ar + $N_2$ |
| **Argon gas flow rate** | 8 sccm |
| **Nitrogen gas flow rate** | 2 sccm |
| **Sputtering pressure** | 0.4 Pa |
| **TiN layer thickness** | 5 nm |
| **AlN layer thickness** | 1 to 5 nm |
| **Number of bilayers** | 234, 200, 175, 156, 140 |
| **Total film thickness** | ~ 1.4 µm |
| **Sputtering power for Ti** | 250 W |
| **Sputtering power for Al** | 250 W |
| **Substrate bias** | 70V |
| **Substrate temperature** | 475 ºC |
| **Substrate** | Silicon (100) |

The TiN layer thickness was kept constant at 5 nm and the AlN layer thickness was varied in steps of 1 nm from 1 to 5 nm thickness, producing five sets of samples: i) 5 nm (TiN) -1 nm (AlN), ii) 5 nm (TiN) - 2 nm (AlN), iii) 5 nm (TiN) - 3 nm (AlN), iv) 5 nm (TiN) - 4 nm (AlN), v) 5 nm (TiN) - 5 nm (AlN). For ease of representation, the samples will be denominated using TiN layer thicknesses as the first number followed by AlN layer thickness as the second number separated by a dash, for e.g. sample 5-3 corresponds to alternating 5 nm thick TiN and 3 nm thick AlN layers. Therefore, the five samples from this study are named as 5-1, 5-2, 5-3, 5-4 and 5-5.

*2.2. Structural characterization details*

Grazing incidence X-ray diffraction (Bruker AXS, D8 Discover) measurements were made using $CuK_{\alpha}$ radiation of wavelength ($\lambda$ = 1.5418 Å) to characterize the TiN-AlN multilayer films. The excitation voltage and current in the diffractometer were set to 40 kV and 40 mA, respectively. The angle of incidence was kept constant at $1^{o}$ while step size of $0.02^{o}$ and scan range from 20 to $90^{o}$ were used.

Transmission electron microscope (TEM) studies were conducted to determine the thickness and structure of TiN-AlN multilayer films. The TEM lamellae were prepared from the selected films (5-1, 5-3 and 5-5) using the standard focused ion beam (FIB) lift-out procedures (Zeiss Crossbeam 540) utilizing $Ga^{+}$ ions at 30 kV. A platinum layer was deposited on the selected region to prevent any ion beam interaction with the film layers. Final polishing of all lamellae was performed at 5 kV and 2 kV to remove any amorphized regions that were developed during the milling procedure. The lamellae were thinned down to an approximate thickness of 80 nm for electron transparency. Structural characterization of the films was performed inside a TEM (Jeol JEM F200) at operating voltage of 200 kV in imaging and diffraction modes.

*2.3. Micromechanical characterization details*

Nanoindentation of all the five TiN-AlN multilayer films was performed using a diamond Berkovich tip in a Hysitron Ubi-1 nanoindenter (Hysitron, Minneapolis, USA). A minimum of 10 indents were performed on each sample at room temperature. Indentations were carried out in load control mode with a maximum applied load of 6 mN, peak hold time of 5 seconds and loading/unloading rate of 0.6 mN per second. The maximum indentation depth in all the samples was less than 100 nm, which is <10 % of film thickness, to avoid substrate effect [26]. This enabled extraction of hardness and indentation modulus of the samples using the commonly used method outlined by Oliver and Pharr [27]. Compliance and tip area calibrations were performed on fused silica samples. In addition to nanoindentation, scratch testing was performed to understand the delamination and fracture properties of the TiN-AlN films. Scratch testing of the five films was performed using a MTS nanoindenter XP using a 10 µm radius cono-spherical diamond tip. Three scratches were performed on each film using a ramped load profile from 0 to 500 mN with loading ramp of 100 mN/sec and scratch speed of 10 μm/s. Scanning electron microscope (SEM) imaging was used for post-mortem analysis of the scratches.

Micropillars were fabricated in the 5-1, 5-3 and 5-5 films by FIB milling using a Tescan Lyra dual beam SEM/FIB workstation utilizing $Ga^+$ ions at 30 kV. Milling of the pillars was carried out with an ion current of 1.2 nA while fine polishing was done at 0.1 nA. The final diameter of the micropillars varied from 0.6 - 0.8 μm with a height/diameter aspect ratio of ~2.5. It is necessary to maintain an aspect ratio less than 3 to avoid buckling. The milled craters were large enough to allow viewing of the entire pillar height, which was typically larger than the film thickness. This resulted in the Si (100) substrate forming a part of the pillars at the base, thus minimizing the sinking-in of the pillars during micro-compression. Micro-compression tests were performed *in situ* in a Zeiss DSM 962 SEM using an Alemnis nanoindenter [28]. A 5 μm diameter diamond flat punch indenter was used for microcompression testing. A minimum of 6 pillars were compressed at room temperature for each of the three films. Engineering stress-strain curves were computed from the load-displacement curves using the top diameter and height measurements from SEM (Hitachi S4800) of the pre-compressed micropillars. Since the films were supposed to show cracking and brittle behavior, the pillars were compressed in displacement-controlled mode using a constant strain rate of 0.001/s taking advantage of the intrinsic displacement control feature of the Alemnis indenter. This avoids the indenter crashing onto the pillars at the onset of cracking during compression (when compared to load-controlled indentation/compression) and allows imaging of the cracked pillars post-deformation. Frame compliance of the Alemnis system was calibrated on standard fused silica samples and corrected from the load-displacement curves. Due to the Si substrate forming a part of the pillar, no elastic sink-in correction was performed while computing the stress-strain curves. All micropillars were imaged after the microcompression tests for post-mortem analysis to identify the deformation behavior.

## 3. Results

### *3.1 XRD analysis:*

**Figure 1** shows the grazing incident X-ray diffraction (GI-XRD) patterns obtained from the multilayer TiN-AlN films deposited with constant 5 nm TiN layer thickness and varying AlN layer thickness from 1 to 5 nm on Si (100) substrates. The 5-1 sample, consisting of 5 nm thick TiN and 1 nm thick AlN layer, shows a strong (111) and (200) texture and both the layers retain cubic structure. The texture effect reduces for the 5-2 sample and the (111) and (200) peaks shift towards higher diffraction angles suggesting a slight decrease in lattice parameter. The small XRD responses on both sides of the (111) and (200) peaks are satellite reflections indicative for superlattice formation. With further increase in AlN layer thickness to 3 nm, a new peak (0002) can be detected at a position associated with the AlN wurtzite phase with hexagonal B4 structure. This suggests that the transition of AlN layer from rock salt cubic structure to B4 hexagonal wurtzite structure occurs somewhere between 2-3 nm layer thickness. This hexagonal peak (0002) gradually shifts to lower diffraction angles with increasing AlN layer thickness from 3 to 5 nm, indicating a decrease in the lattice parameter value, which is in proportion to the fraction of the Al content in the film.

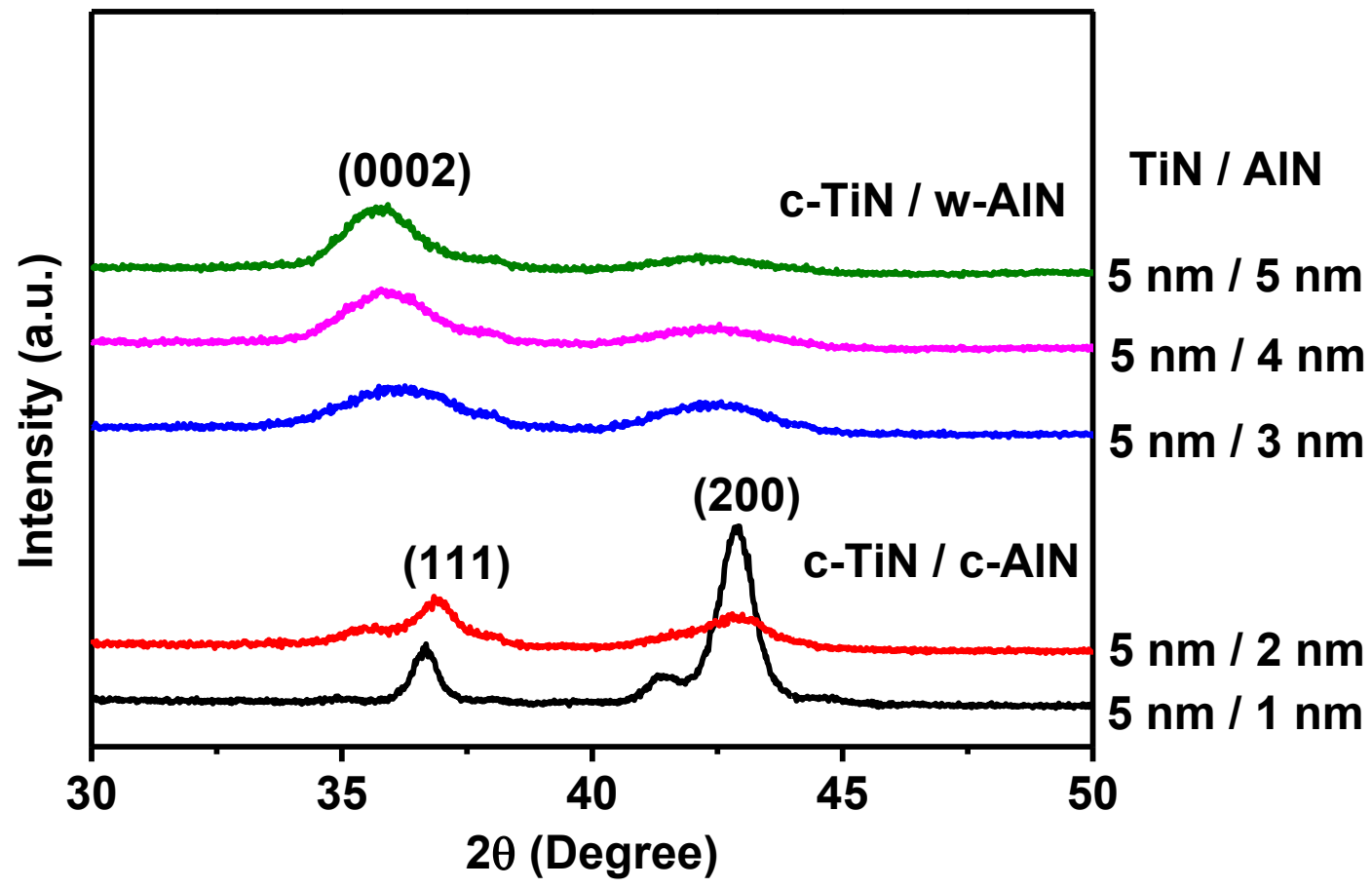

**Figure 1:** GI-XRD spectrum of TiN-AlN multilayer films with varying AlN layer thickness from 1 to 5 nm.

*3.2 TEM studies*

**Figure 2(a)** shows the TEM image of the samples holding 5 nm thin TiN with 1, 3, 5 nm thin AlN layers. TEM studies confirm the thickness of the individual layers of the TiN and AlN, with homogeneous layer thickness in all the films. The TiN and AlN interface in the 5-1 sample appear to be straight and sharp whereas in the 5-3 and 5-5 samples, they appear wavy. The selected area electron diffraction (SAED) from the films confirms the crystalline structure of both layers in all films, as shown in **Figure 2(b)**. In the case of the 5-1 films, the AlN lattice planes are well ordered having perfect lattice matching with the atomic arrangements of TiN forming a superlattice structure (Figure 2(a) inset). The electron diffraction pattern of this 5-1 sample is more spotty (or a discontinuous ring pattern) and is indexed as rock salt type cubic structure for both TiN and AlN layers, growing with a strong (111) and (200) texture. The 5-3 and 5-5 films show textured and more continuous ring pattern of nanocrystalline structures with hexagonal wurtzite structure from AlN. The w-AlN patterns are indexed as $(10\bar{1}0)$ and (0002) planes, with a strong (0002) texture. The TEM results are well aligned to the XRD observation confirming the superlattice structure in 5-1 films and the phase transition of AlN from cubic to wurtzite for 5-3 and 5-5 films.

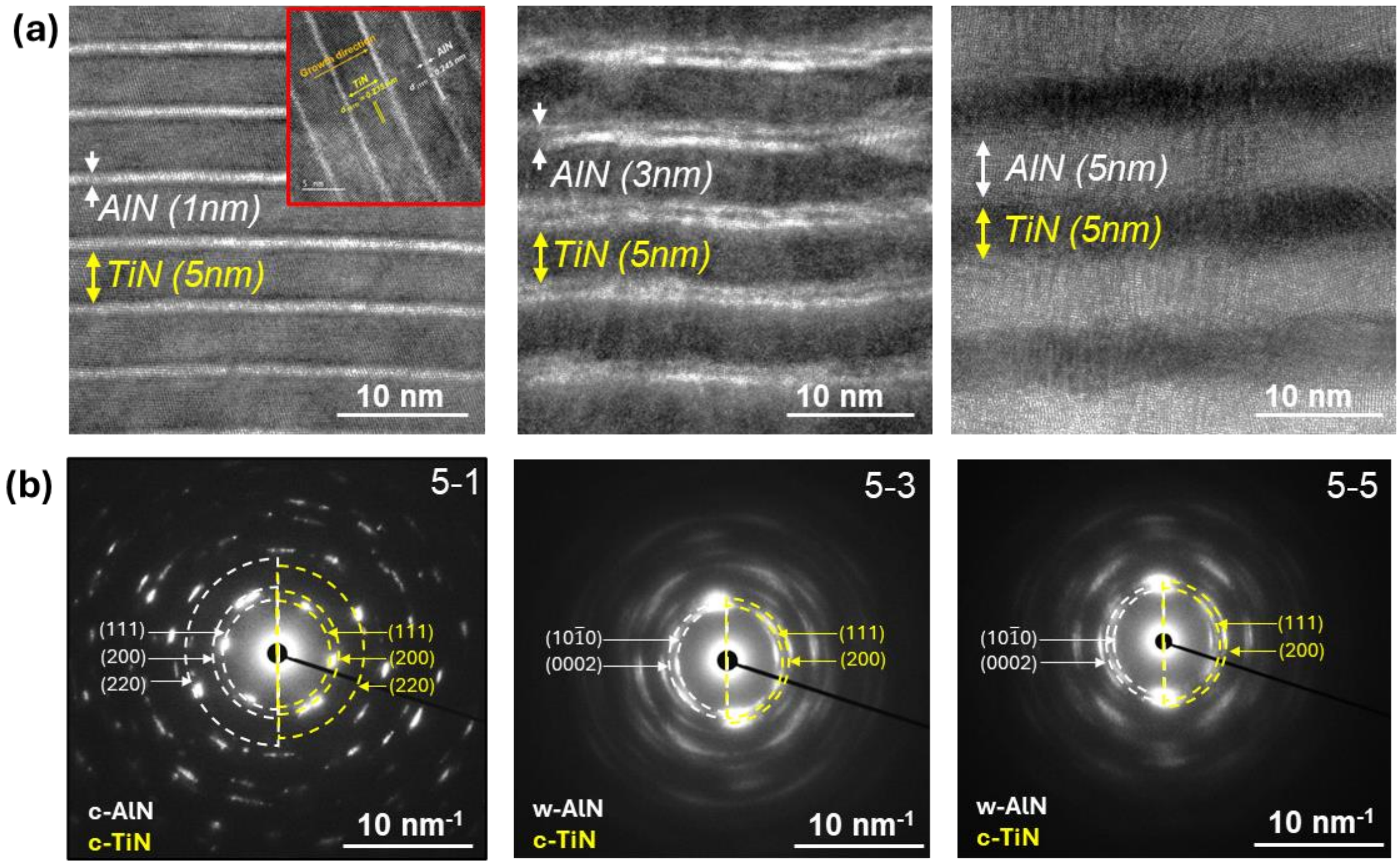

**Figure 2: (a)** TEM micrographs and **(b)** SAED diffraction patterns of TiN-AlN multilayer films with 5 nm thin TiN layers combined with either 1 nm thin, 3 nm thin, or 5 nm thin AlN layers. Figure 2a inset shows the HR-TEM image of 5-1 films showing the coherent interface.

*3.2 Nanoindentation results*

Nanoindentation provides an initial estimate of the elastic and plastic properties of the multilayer TiN-AlN films. **Figure 3(a)** shows the representative load-displacement curves for these films and **Figure 3(b)** shows their hardness (*H*) and indentation modulus (*E*) values as a function of their AlN layer thickness (or the sample tested). The 5-1 sample has the highest hardness and indentation modulus values of 33 GPa and 325 GPa, respectively among all the films. The measured hardness value is around 38 % higher than the hardness value of single layer TiN film (*H*= 24 GPa) [29]. Hardness and indentation modulus values decrease with increasing AlN layer thickness eventually plateauing out for AlN layer thicknesses of 4 and 5 nm. The drop in mechanical properties with increasing AlN content suggests that the superlattice effect reached its maximum (for our layer configurations) with the 5-1 sample. The transition of AlN from cubic to hexagonal (for AlN layer thicknesses between 2 to 3 nm) is accompanied by a steeper drop in hardness and indentation modulus values, in comparison to the change observed from 5-1 to 5-2 samples. This further proves that hexagonal AlN has lower hardness and indentation modulus in comparison to its cubic counterpart. We must note that the total numbers of layers interrogated in nanoindentation does not vary a lot between 5-1 and 5-5 samples to affect these results. If we consider the plastic deformation volume and estimate the plastic depth to be ~ 3x the indentation depth, the number of layers investigated for 5-1 and 5-5 samples are 40 and 29, respectively. So, the change in nanoindentation hardness and indentation modulus observed in these samples is a real property of the composite, and not an artefact of measurement or data analysis. No obvious pop-ins or serrations, which are typical characteristics of cracking, were observed in the load-displacement curves for all the five samples. It should be noted that for such shallow indentation depths (< 100 nm in this case), the hardness values are typically higher than the bulk property measurements due to indentation size effect. In addition, the tip area calibration for nanoindentation corrects the indentation modulus and hardness values. Nevertheless, for the same indentation parameters used on both samples, the results can be compared.

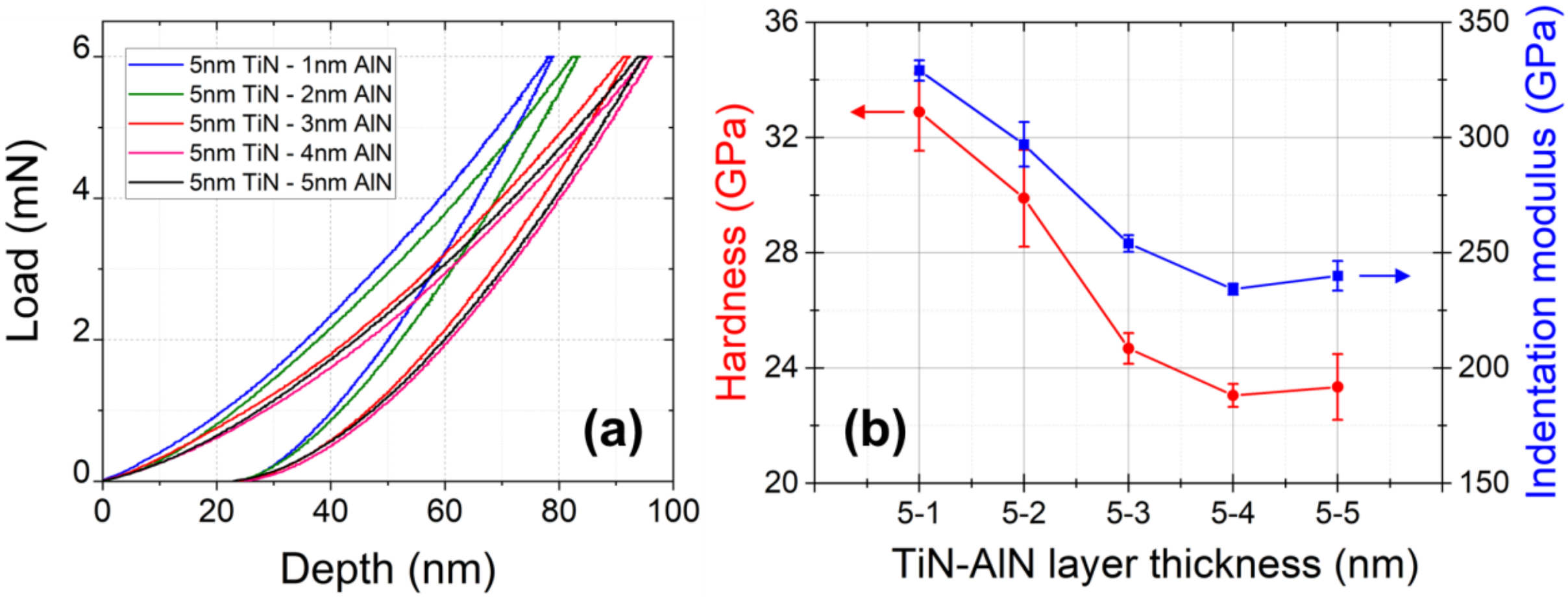

**Figure 3: (a)** Nanoindentation load-displacement curves and **(b)** hardness and indentation modulus for TiN-AlN multilayer films with varying AlN layer thickness from 1 to 5 nm.

*3.3 Scratch test*

Ramped load, lateral scratch tests are typically used to qualitatively assess the adhesion of films to substrates. These tests are influenced by both intrinsic factors pertaining to the film-substrate system being tested and extrinsic factors related to the test methodology, that precludes a more quantitative measurement of film adhesion. The purpose of the nanoindentation scratch tests reported in this paper is to compare the delamination loads for each film for constant experimental parameters and to look at changes in cracking behavior of the films as a function of increasing AlN layer thickness. The ramped load scratch tests showed film delamination for all films except the 5-3 multilayer system (**Figure 4**), which was quite unexpected. The scratch load at which delamination was first observed was found to vary strongly between the films, as indicated in **Figure 4**. The 5-1 sample exhibited the lowest delamination load at 160 ± 10 mN while the 5-5 sample was observed to have the highest at 320 ± 10 mN. The delamination loads for the 5-2 and 5-4 samples lie between these two extremes at 210 ± 30 mN and 300 ± 70 mN respectively. The 5-3 film, however, did not show any delamination up to the maximum tested load of 500 mN. It did show microcracking along the scratch path, like other films.

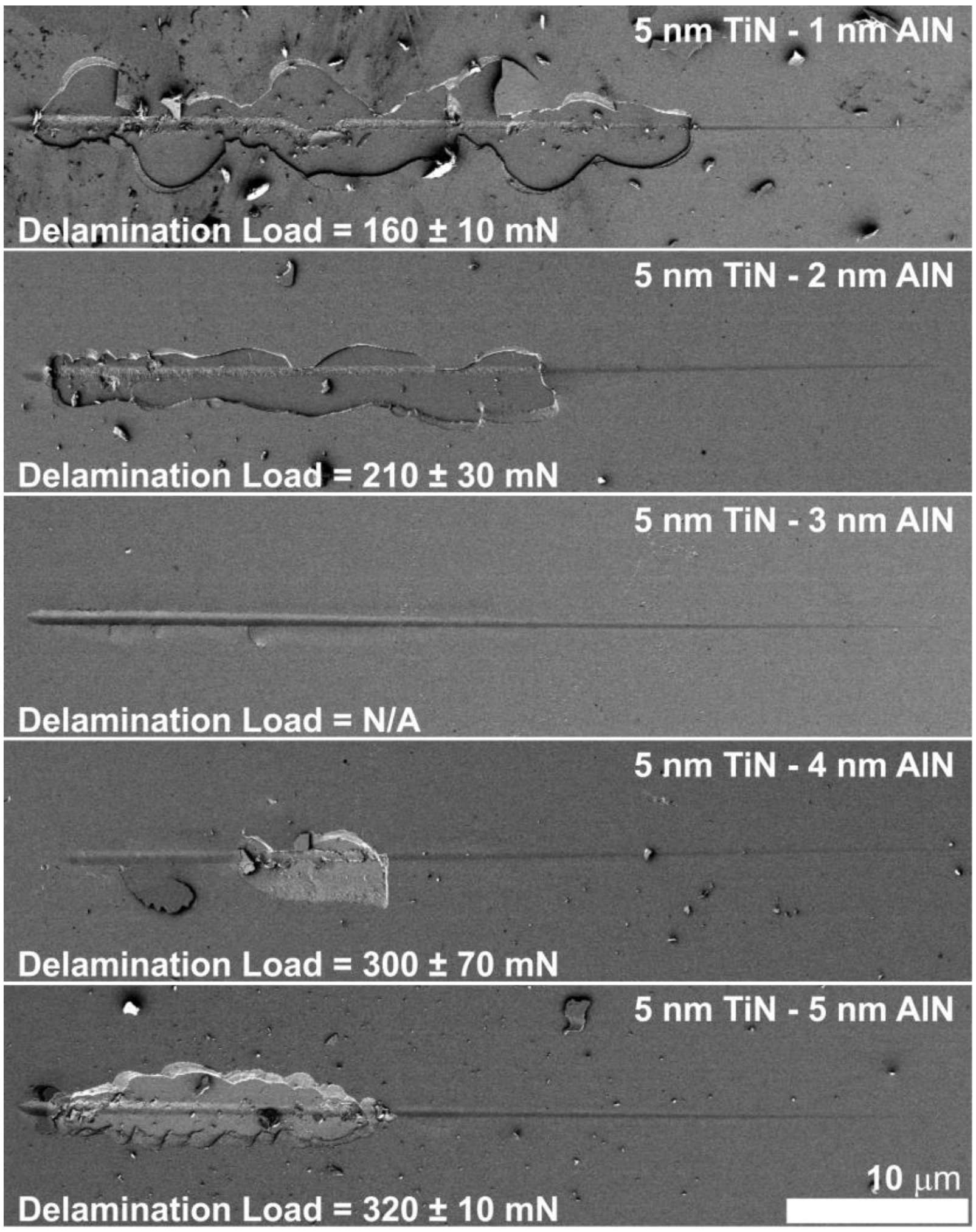

**Figure 4:** SEM images of scratches ramped from 0 mN to 500 mN in the five multilayer films. The average and standard deviation of load for the onset of delamination of the film is shown on the bottom left of each image. The lack of delamination in the 5 nm TiN - 3 nm AlN multilayer is notable.

Cracking of the films along the scratch path was observed in all cases with the onset of cracking at around 250 mN. The fracture and delamination behavior of these multilayer films can be divided into two distinct regimes: (a) fracture of film immediately followed by delamination of film (samples 5-1 and 5-2), and (b) fracture of film with microcracking, resulting in higher loads required for delamination (samples 5-3, 5-4, and 5-5). These regimes are further illustrated in **Figure 5** with higher magnification images of the cracks developed along the scratch path. **Figure 5(a)** shows the fracture surface of the 5-2 sample; it can be seen that the fractured edge of the film is relatively smooth and vertical to the film growth direction with minimal crack deflection along the layer interfaces. The fracture surface of the 5-1 sample is also similar. **Figure 5(b)** shows the scratch path in the 5-3 film. The indenter tip initiates many cracks in the film but the dissipation of elastic strain energy through microcracking prevents film delamination. **Figure 5(c)** shows the fracture surface from the 5-4 sample, which shows similar fracture behavior as the 5-5 film. The evidence of extensive microcracking can be seen through the presence of many small delaminations on the fracture surface. The 5-4 and 5-5 films mostly fracture along a path inclined to the substrate, when compared to the 5-1 and 5-2 samples that show a vertical fracture path towards the substrate (refer **Figures 4 and 5**). Inclined fracture paths increase the fracture toughness of the film due to increase in newly created surfaces. Cracks developed in the films during the scratch test provide a qualitative idea about the crack propagation behavior and fracture toughness in these films. The 5-3 film seems to exhibit the highest fracture resistance among all the investigated films as it did not delaminate up to the maximum applied load of 500 mN and shows extensive microcracking. This effect seems to decrease as we move to 5-4 film which shows a very interesting delamination behavior. Delamination was seen to halt after initiation in two of the three scratches performed; one of these scratches is shown in **Figure 4**. This is very unusual in ramped loading scratch tests where the mechanical driving force for film delamination continues to increase as the scratch progresses.

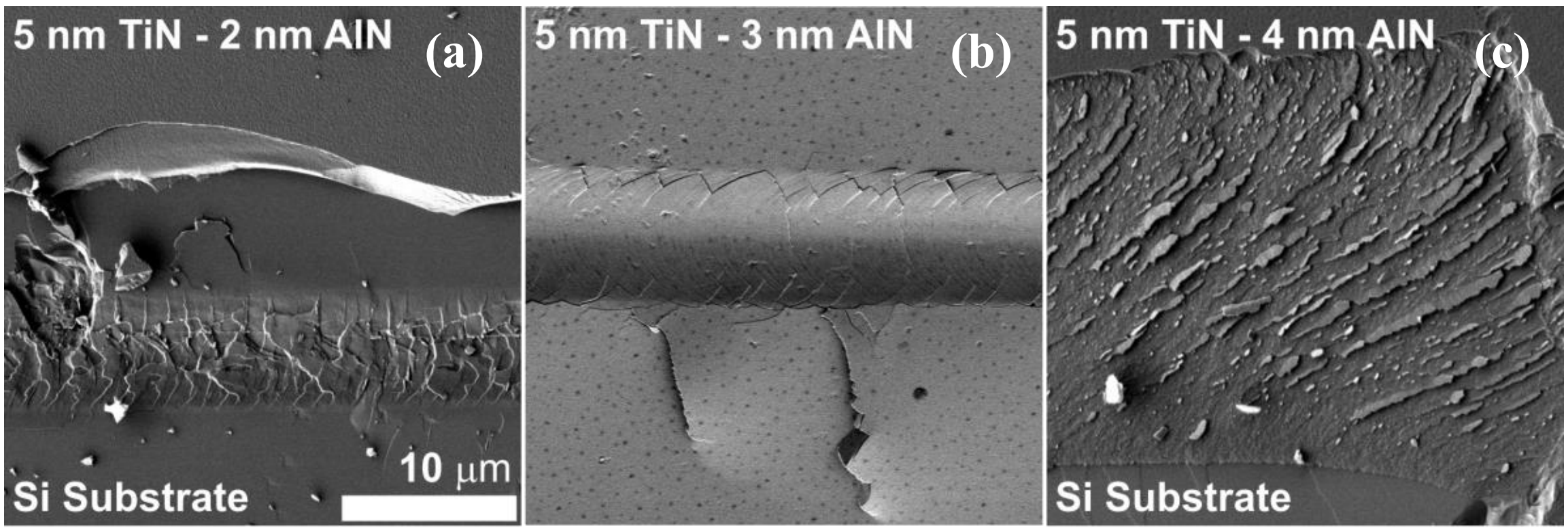

**Figure 5:** Higher magnification images of the film scratch surfaces: **(a)** 5 nm – 2 nm, **(b)** 5 nm - 3 nm, and **(c)** 5 nm – 4 nm. Note the high density of cracks in the 5 nm – 3 nm film and the evidence of interfacial microcracks in the 5 nm – 4 nm film.

*3.4 Microcompression test*

Due to slight differences in film thickness and FIB milling behavior of the films, the Si substrate forms a part of the pillar base. It comprises about 36 % for 5-1, 22 % for 5-3 and 13 % for 5-5 sample by height thereby, influencing the yield strength and flow stress values of the compressed micropillars. For this reason, the microcompression data between different samples will not be compared directly, unlike the nanoindentation data. They should be considered for a more qualitative analysis to compare the compression behavior of these multilayer films.

The engineering stress-strain curves for the most consistent set of data obtained from 5-1, 5-3 and 5-5 films are shown in **Figure 6** along with the post compression SEM images at two different strains. The micropillars exhibit roughly similar yield strengths (~15 GPa) although the Si substrate forms different volume fractions of the pillars for the three samples. The pillars were compressed to different strains to understand their failure and deformation behavior. Beyond the yield strength, the pillars exhibit decreasing stresses with increasing strain due to onset of cracking. This is a common feature in all three samples; however, they show different stress drop and cracking characteristics. For instance, sample 5-1 shows a gradual or "stepped" stress decrease consistent with the progressive formation of lots of small cracks. The post deformation images of the 5-1 pillars show extremely brittle failure behavior with lots of cracks and formation of many small fragments in the pillars. Due to brittle behavior, the stress values remain abysmally low at higher strains (> 0.15) as these small fragments do not support the load exerted by the flat punch indenter. On the contrary, sample 5-3 shows a rapid stress drop that can be clearly correlated with slightly ductile shear fracture in the pillars. The post compression images show nice cleavage planes forming at an angle to the loading direction, suggesting extensive crack deflection at the multilayer interfaces. On further loading, the stress starts increasing slightly due to plastic deformation of the top region of the pillar. The post compression image of a 5-3 pillar deformed to 0.2 strain shows substantial plastic deformation at the top. Sample 5-5 exhibits mechanical behavior closer to 5-3 than 5-1 sample. Nevertheless, there are slight differences. The stress drops are smaller in magnitude in comparison to 5-3, suggesting greater crack deflection at the interfaces and possible crack bridging. The stress rises steadily with increasing strain beyond the first stress drop which can be correlated with steady plastic deformation of the pillars. The SEM image of a pillar compressed to 0.35 strain shows extensive plastic deformation. Performing the *in-situ* micropillar compression in the SEM allows correlation of the stress-strain curves directly with the pillar deformation characteristics.

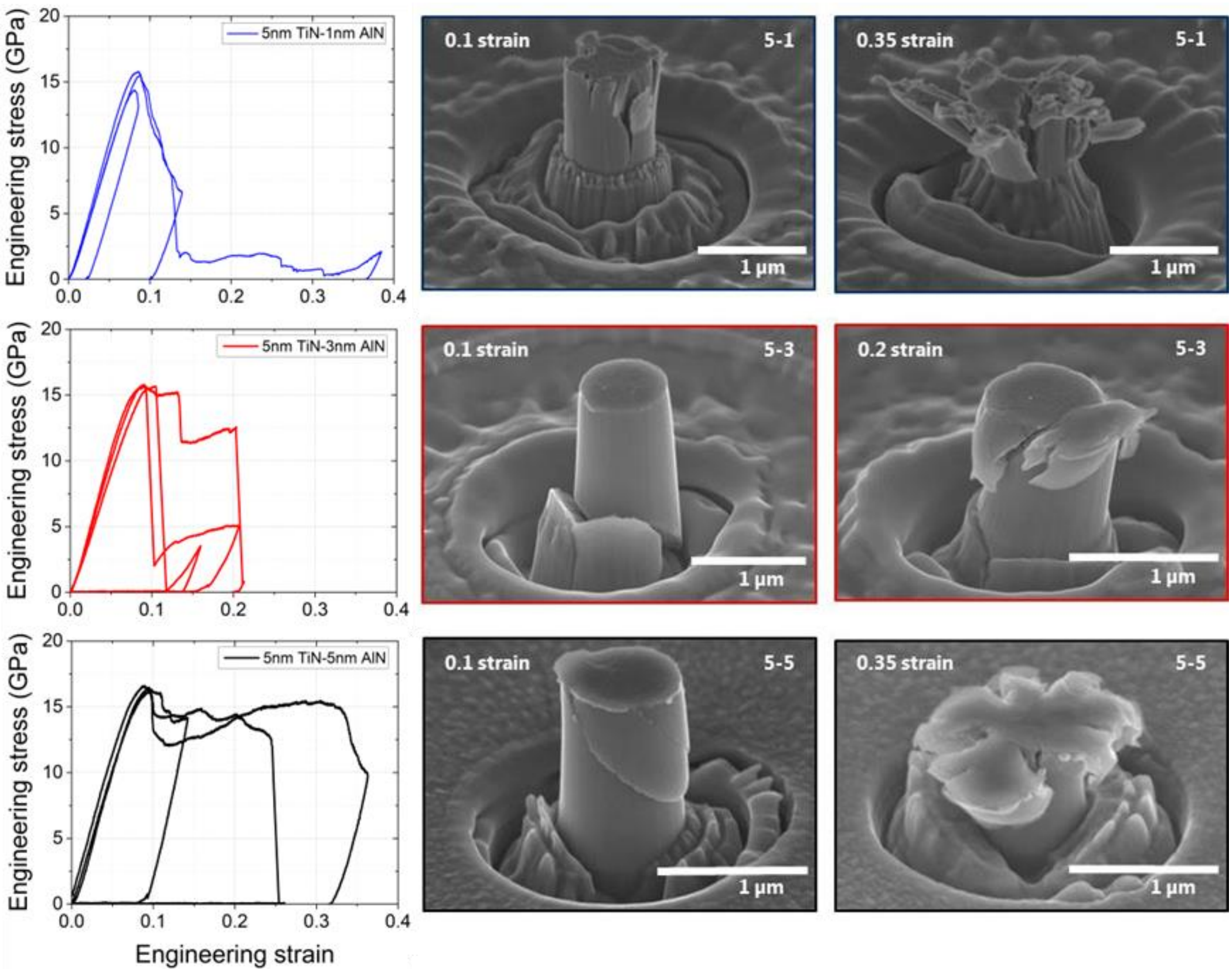

**Figure 6:** Micropillar compression results for samples 5-1, 5-3 and 5-5. On the left, engineering stress-strain curves are plotted. On the right, SEM images of the pillars post-compression are shown for two different strains to underline the failure behavior of these multilayers.

## 4. Discussion

The X-ray diffractograms for the five films indicate that a structural shift takes place in the films as the thickness of the AlN layer is increased: from c-AlN in the 5-1 and 5-2 samples to w-AlN in the 5-3, 5-4 and 5-5 samples. From literature [11,12,30,19,24,31,32] of TiN-AlN systems, it is evident that c-AlN has only ~ 4 % lattice difference with c-TiN [33], hence, metastable c-AlN films with a critical thickness of 2–3 nm could be stabilized on TiN templates. Beyond the said range, stable hexagonal or mixed hexagonal and cubic structure of AlN phase is observed, which depends highly on the growth conditions of the films. Chawla et al. [34] studied and predicted the equilibrium critical thickness up to which the cubic AlN phase is energetically favored to the wurtzite variant in TiN-AlN systems in detail and concluded that to stabilize c-AlN by growing on TiN, the mechanical properties of the substrate, the crystallographic orientation and lattice parameters, as well as the coherency state of the substrate–layer interface play a crucial role. The phase change from c-AlN to w-AlN is evident from the TEM diffraction patterns which shows the strong orientation towards $(10\bar{1}0)$ and (0002) peaks of wurtzite phase for AlN layer thickness of 3 nm and above [8]. The mechanical characterization of the films using nanoindentation, scratch and micropillar testing indicates that this structural shift is accompanied by a shift in the mechanical performance of the multilayers. Setoyama et al. [8] and Madan et. al [15] showed the epitaxial relationship of TiN-AlN interface and the existence of c-TiN and c-AlN superlattices with TiN//AlN(001) relationship for AlN thickness less than 1.5 nm. This transforms to a stable hexagonal w-AlN phase with (111)TiN//(0001)AlN orientation relationship for larger AlN thicknesses. The increase in thickness of the film layers results in the breakdown of the cubic superlattice which has direct implication on the mechanical properties. There is around 36 % reduction in the hardness of the system due to the breakdown of the cubic superlattice. These results are in good agreement with the results of Kim et al. [11] which show that the hardness decreases from 33 GPa to 21 GPa when the period thickness was reduced from 5 nm to 1 nm. This confirms the effect of superlattice formation and elastic modulus mismatch between the c-TiN and c-AlN on the improvement in the overall mechanical properties in terms of hardness.

The mechanical characterization of the films using nanoindentation, scratch and micropillar testing indicates that this structural shift is accompanied by a shift in the mechanical performance of the multilayers. The hardness values decreases with increase in the AlN layer thickness and the maximum value of hardness is observed for the multilayer films when the multilayer forms a superlattice [11]. For the layers in this study, this is the 5-1 arrangement, which has a significantly higher hardness than the TiN film itself. The so called superlattice effect – due to the high shear modulus difference and coherent interface between the cubic TiN and AlN layers [11,35–38] is most pronounced for the 5-1 film. A further increase of the AlN layer thickness leads to a decrease in hardness and especially if the critical layer thickness is exceeded, the formation of the thermodynamically stable h-AlN is preferred (as the coherency strains with TiN only allow to stabilize AlN in its metastable cubic structure up to about 2–3 nm) leading to an even steeper decline in hardness and indentation modulus. The decrease in hardness and indentation modulus of our multilayers as soon as w-AlN is formed, is certainly to be expected as w-AlN ($H$ = 17 GPa and $E$ = 270 GPa) [39–43] is both softer and less stiff than TiN ($H$ = 24 GPa and $E$ = 320 GPa) [44–46] and c-AlN ($E$ = 501 GPa) [34].

In micropillar compression tests, the 5-1 sample is observed to behave in a very brittle manner while the 5-3 and 5-5 samples exhibit cracking and re-loading. The complications involved in the micropillar compression of a 1.4 μm thick hard film mean that it is not possible to make useful comparisons between the 5-3 and 5-5 films. However, the instantaneous fracture of the 5-1 film compared to the cracking and re-loading of the 5-3 and 5-5 films is further evidence of increased toughness through crack stopping in these latter multilayers. In sample 5-1, both TiN and AlN layers are in cubic phase forming a superlattice structure with coherent interface where the AlN layer having thickness of 1 nm is inadequate to interrupt and arrest the movement of the crack between the TiN layers. The cracks can originate from the column boundaries resulting from the columnar growth, as the columnar growth is not interrupted due to coherent growth of c-AlN over TiN layers. Hence, the formation of lots of small cracks were observed in the micropillar sample 5-1 followed by instantaneous fracture as coherent interfaces do not deflect cracks. On the contrary, samples 5-3 and 5-5, which have 3 nm and 5 nm thick AlN layers, respectively, i.e. w-AlN phases and no superlattice is formed. This is the reason that sample 5-5 exhibits a mechanical behavior closer to the 5-3 than to 5-1 sample. Due to the thick AlN layers - with only partial coherent interfaces - in both 5-3 and 5-5 samples, AlN layers were able to interrupt the movement of the crack at the multilayer interfaces. The stabilization of wurtzite AlN phase improved the toughness of the system as seen from the micropillar compression results of 5-5 samples, where the pillars were deformed up to 0.35 strain. This is mainly due to the lower indentation modulus and hardness of the w-AlN compared to the c-AlN where the crack generated in the brittle TiN layer gets probably deflected and absorbed by the less stiff layer.

The increased toughness of the multilayers incorporating w-AlN, as demonstrated by micropillar compression and scratch testing, is a far more important result. Of particular interest is that the 5-3 film exhibits the highest fracture resistance in a scratch geometry. In the scratch test geometry, the load at which delamination begins and the extent to which the film delaminates to the sides of the scratch can be used to assess the extrinsic adhesion of the film. In the present case, the interface between the multilayer film and the silicon substrate remains the same and hence, changes in scratch performance must be related to energy dissipation mechanisms within the multilayers rather than the intrinsic adhesion at the interface to the substrate. In the "brittle" regime, seen in **Figure 5(a)** for the 5-1 and 5-2 multilayers, the inclusion of more AlN increases the delamination load and decreases the extent of film delamination, **Figure 4**. This is a result of the driving force for delamination, i.e. the elastic strain energy, being reduced for a given load. The nanoindentation data, **Figure 3**, shows that the indentation modulus of the 5-2 multilayer is around 10 % lower than that of the 5-1 film. In the "microcracking" regime, seen in **Figure 5(b) and 5(c)**, the inclusion of thicker AlN layers decreases the delamination load and increases the extent of film delamination. The nanoindentation data, **Figure 3**, indicates that the elastic moduli of the 5-3, 5-4 and 5-5 multilayer films are very similar and hence the elastic driving force for delamination is also similar. However, in these films, it seems that extensive microcracking is the cause of the higher delamination loads and reduced extent of delamination. In fact, the toughening of the 5-3 film was so successful that scratching at the highest available load did not cause delamination in any of the scratches performed. The higher fracture resistance of the 5-3 film compared to the 5-4 and 5-5 films can be explained if it is assumed that the observed microcracking is due to the relatively weak interface between the c-TiN and w-AlN layers. It has been demonstrated that toughening via interfacial crack deflection requires the fracture toughness of the interfaces to be rather lower than that of the matrix [47,48]. In this case, the density of these weak interfaces will principally affect the observed toughness – a higher volume of weak interfaces results in a higher density of microcracks and more energy absorption. In the present case, the 5-5 film contains approximately 20 % fewer c-TiN/w-AlN interfaces than the 5-3 film, for the same thickness. Thus, in TiN-AlN multilayer systems, the cubic to tetragonal phase transformation is tuned by the nanoscale layer design to improve structural integrity of these systems in critical extreme environment applications such as cutting tool industry and high-temperature resistant coatings.

## 5. Conclusion

TiN-AlN multilayer film systems with constant TiN thickness of 5 nm and varying AlN layer thicknesses from 1 to 5 nm were grown by the reactive magnetron sputtering. The XRD results showed that cubic phases existed in both TiN and AlN for the AlN thickness up to 2 nm, whereas cubic phase transformed into hexagonal wurtzite phase for thicker AlN layers. This phase transformation was accompanied by changes in the mechanical properties of the film systems. Hardness and indentation modulus values decreased with increase in the AlN layer thickness, with films containing the c-AlN phase having better mechanical properties compared to those with the w-AlN phase. The 5-1 multilayer exhibits the most pronounced superlattice effect in hardness with 33 GPa and indentation modulus of 330 GPa. There is approximately 30 % reduction in hardness and 27 % reduction in elastic modulus associated with the formation of w-AlN, upon increasing the AlN layer thickness to 5 nm. The micro-compression test results obtained yield strengths (~15 GPa) for all the samples where brittle to ductile transition was observed with increase in AlN thickness. Here, the 5-3 sample (5 nm thin TiN and 3 nm thin AlN layers) exhibits the highest fracture resistance, as there is no instantaneous brittle fracture (as obtained for the 5-1 sample), rather an interrupted crack growth, indicating mechanisms that even allow for plasticity-related effects. Also, the 5-5 sample shows a similar behavior during the compression test but to a less-pronounced extent. The scratch test resulted in an increased crack resistance in the 3 nm thin AlN film containing TiN-AlN as the crack deflection along the interface reduced the overall cracking of the films. These results clearly show that the phase transformation of c-AlN to w-AlN has a significant effect on mechanical properties of the TiN-AlN multilayers. The superlattice and multilayer concept in TiN-AlN films can thus be used in demanding applications where extreme surface performance is required or in harsh working environments.

**Acknowledgments**

N.G.M and G.M acknowledge funding from Research Council of Finland (RCF) through HERBIE project (funding decision 341050). V.C. acknowledges financial support from the LISE MEITNER Program (Project No M1233) of the Austrian Science Fund (FWF), FP7: People and Marie-Curie action COFUND fellowship under EMPAPOSTDOC Cofund program. This work made use of H2MIRI (funded through RCF decision number 353235) and Tampere Microscopy Centre facilities at Tampere University.